\documentclass[cameraready]{Interspeech}

\title{\textsc{CtrlSpeech}: Coarse-to-Fine Control for Expressive Speech Synthesis}

\author[affiliation={1}]{Zhisheng}{Zheng}
\author[affiliation={2}]{Xiaohang}{Sun}
\author[affiliation={2}]{Zhu}{Liu}
\author[affiliation={2}]{Caren}{Chen}
\author[affiliation={2}]{Rohith}{Kumar}
\author[affiliation={2}]{Manoj}{Aggarwal}
\author[affiliation={2}]{Gerard}{Medioni}
\author[affiliation={1}]{David}{Harwath}

\address{
    $^1$ The University of Texas at Austin 
    $^2$ Amazon 
}

\email{zszheng@utexas.edu, harwath@utexas.edu}


\newcommand{\modelname}{\textsc{CtrlSpeech}}

\usepackage{comment}
\usepackage{multirow}
\usepackage{amsmath}
\usepackage{amssymb}
\usepackage{bm}
\usepackage{makecell}
\usepackage{booktabs}
\usepackage{graphicx}
\usepackage{xcolor}
\usepackage{pifont}
\usepackage{wasysym}
\usepackage{arydshln}
\usepackage{cite}

\definecolor{SciGreen}{RGB}{44, 160, 44}      
\definecolor{SciRed}{RGB}{214, 39, 40}       
\definecolor{SciOrange}{RGB}{230, 159, 0}

\definecolor{projectblue}{RGB}{0,90,160}

\begin{document}
\maketitle
\begin{abstract}
Recent Text-To-Speech (TTS) systems have achieved strong naturalness and zero-shot voice cloning performance, but fine-grained control of expressive speech at the word or phoneme level remains challenging. We propose \textsc{CtrlSpeech}, a controllable, expressive TTS framework with coarse-to-fine control. Built on the DiTAR architecture, \textsc{CtrlSpeech} combines global speaker conditioning with phone-aligned pitch, loudness, and duration signals, enabling localized prosodic control while preserving the target speaker’s timbre. This design allows users to adjust expressive attributes at a fine temporal granularity, making speech refinement more flexible and controllable. Experimental results show that \textsc{CtrlSpeech} achieves competitive zero-shot TTS performance and improves controllability over expressive attributes, demonstrating its effectiveness for flexible and practical expressive speech synthesis. Our demo, code and model weights are available at \href{https://zhishengzheng.com/ctrlspeech}
{\textcolor{projectblue}{\textbf{\textsc{CtrlSpeech}}}}.
\end{abstract}

\section{Introduction}
Recent text-to-speech (TTS) systems have achieved remarkable progress in naturalness and zero-shot voice cloning. Reference-conditioned methods can synthesize speech for unseen speakers from a short prompt waveform, while text-prompted and instruction-based systems provide a more intuitive interface for controlling speaking style. However, fine-grained and explicit control over expressive speech remains difficult. In many existing systems, speaker identity, prosody, and style are highly entangled. Even when models expose control over multiple speech attributes, such control is often provided only at the sentence or utterance level (e.g., a global style prompt or instruction), rather than as temporally aligned signals tied to specific phones or words. As a result, users still lack a practical way to precisely manipulate local prosodic events—such as the pitch, loudness, or duration of a specific segment—while preserving target timbre.

To address this problem, we propose \modelname{}, a controllable, expressive speech synthesis framework with \emph{coarse-to-fine} control. Built on a DiTAR~\cite{jia2025ditar} backbone, \modelname{} combines global speaker conditions for timbre control with phone-aligned prosodic signals, including pitch, loudness, and duration, enabling explicit manipulation of local expressiveness while preserving speaker identity. It also supports iterative refinement, allowing users to progressively improve synthesized speech through control adjustment, which makes expressive TTS more practical for real-world use.

The contributions of this work are summarized as follows:
\begin{enumerate}
    \item We propose \modelname{}, a controllable, expressive speech synthesis framework that enables unified coarse-to-fine control over both global and local attributes.
    \item We design an explicit control pipeline that supports iterative speech refinement through aligned pitch, loudness, and duration signals, allowing users to flexibly adjust synthesized speech.
    \item Extensive experiments show that \modelname{} achieves competitive zero-shot TTS performance and significantly improves fine-grained controllability over expressive attributes.
\end{enumerate}



\begin{figure}[t]
    \centering
    \includegraphics[width=1.0\linewidth]{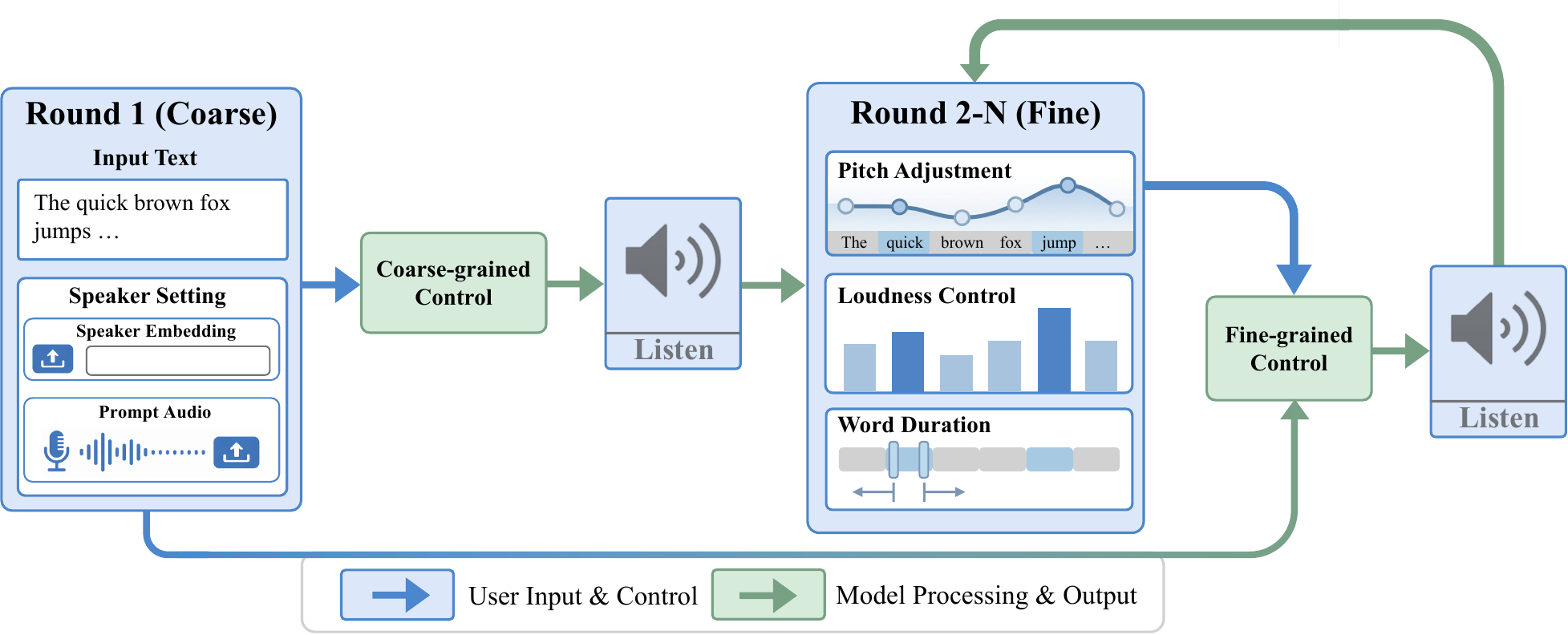}
    \caption{Coarse-to-fine speech control pipeline.}
    \label{fig:control-pipeline}
\end{figure}

\section{Related Work}
\subsection{Reference-Conditioned Speech Synthesis}
Reference-conditioned speech synthesis, also known as voice-cloning or zero-shot TTS, aims to generate target speech by conditioning the model on a short reference utterance from an unseen speaker. The input text specifies the linguistic content and the reference speech provides speaker-dependent cues such as timbre, prosody, emotion, and style. Recent work~\cite{wang2023neural, ju2024naturalspeech, wang2024maskgct, chen2025f5, zheng2025voicecraft, jia2025ditar} shows a clear trend from early style-transfer methods toward more powerful voice cloning and controllable generation: autoregressive LLM-based systems such as VALL-E~\cite{wang2023neural} model speech as discrete codec tokens and enable strong in-context voice imitation, while newer non-autoregressive, diffusion, and flow-based models such as NaturalSpeech~\cite{ju2024naturalspeech}, MaskGCT~\cite{wang2024maskgct}, and F5-TTS~\cite{chen2025f5} improve synthesis quality and efficiency. This paradigm has enabled models to be highly personalized without speaker-specific fine-tuning.

\subsection{Text-Prompted Speech Synthesis}
Text-prompted speech synthesis uses textual descriptions or natural language instructions to guide speech generation in a more intuitive and user-friendly manner. Representative systems~\cite{guo2023prompttts, leng2023prompttts, yang2024instructtts, lyth2024natural} show that language prompts can provide fine-grained control over multiple speech attributes. More recently, LLM-based systems~\cite{zhou2024voxinstruct, yang2025emovoice, diwan2025scaling, du2025cosyvoice, zhang2025vevo2} have further extended this paradigm, enabling more unified and expressive instruction-following speech synthesis.

\section{Methodology}

\begin{figure}[t]
    \centering
    \includegraphics[width=1.0\linewidth]{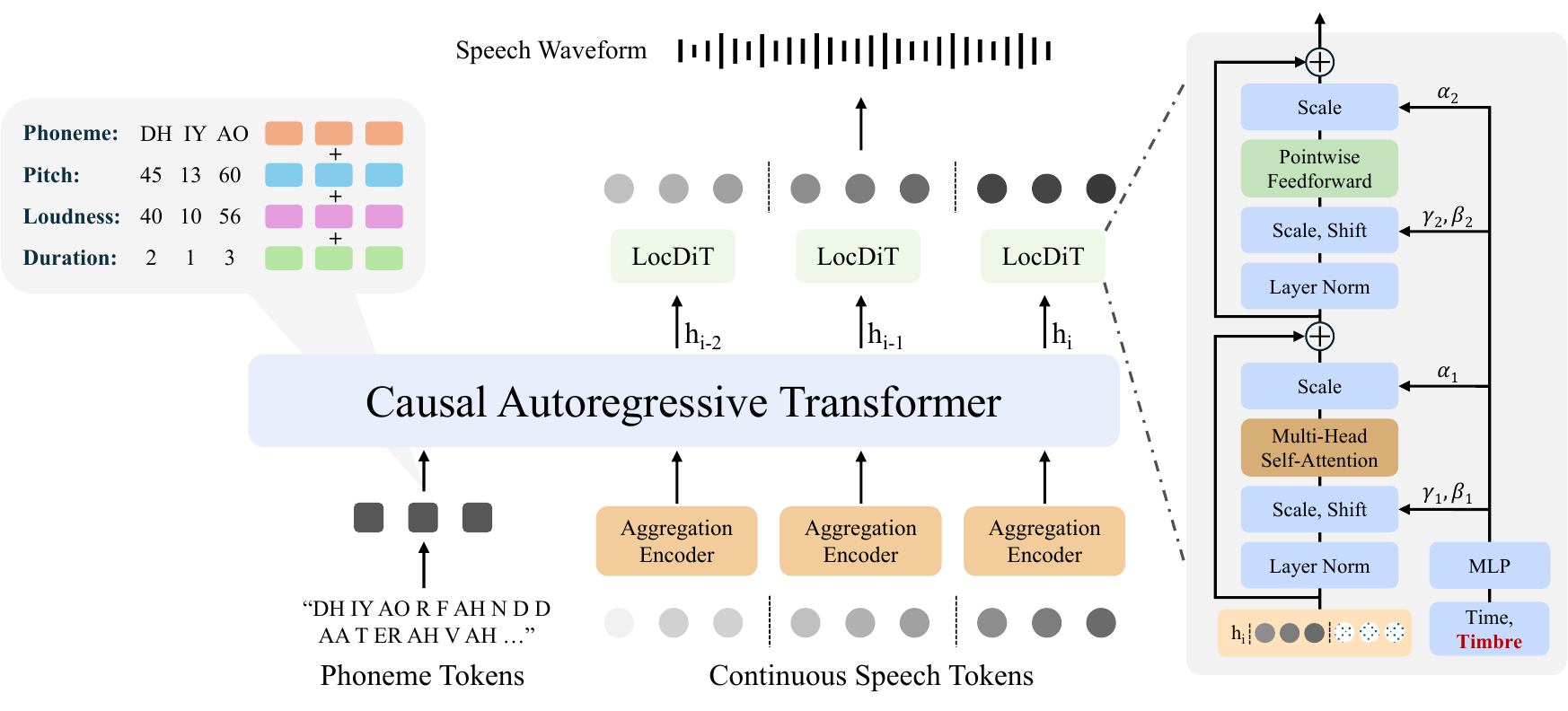}
    \caption{The architecture of \modelname.}
    \label{fig:architecture}
\end{figure}

\subsection{Continuous Speech Representations}
We model speech in a continuous latent space rather than with discrete codec tokens, as continuous representations avoid the information loss introduced by quantization~\cite{jia2025ditar,an2025mela,wang2025streammel}. Prior work also notes that discrete-token pipelines often rely on multi-stage coarse-to-fine generation, which can increase system complexity and accumulate errors~\cite{jia2025ditar,an2025mela}. Therefore, continuous speech representations provide a more suitable space for modeling subtle variations in prosody, timbre, and other fine-grained acoustic attributes.

We adopt a variational auto-encoder (VAE)~\cite{kumar2023high} to transform raw waveforms into compact continuous latent tokens. Specifically, the VAE encoder maps an input waveform to a latent posterior distribution parameterized by $(\mu, \log \sigma^2)$, from which latent tokens are sampled. The encoder is implemented with stacked convolutional layers, while the decoder follows the BigVGAN~\cite{lee2022bigvgan} architecture for waveform reconstruction. In our setup, a 16\,kHz input waveform is compressed into a latent token sequence with a frame rate of 40\,Hz, with each latent token having a dimensionality of 64.

\subsection{Diffusion Transformer Autoregressive Modeling}
\label{sec:ditar_backbone}

To model continuous speech representations, we adopt the DiTAR architecture~\cite{jia2025ditar} as our backbone. DiTAR combines causal autoregressive modeling across patches (groups of tokens) with bidirectional diffusion modeling within each patch. Instead of predicting every continuous token derived from the VAE with a purely left-to-right decoder, we partition the continuous sequence into local groups of tokens (patches) and decouple generation into two levels: an autoregressive module captures long-range dependencies among patches, while a local diffusion transformer reconstructs the fine-grained tokens of the next patch conditioned on the autoregressive state. This design leverages bidirectional attention to better model short-range correlations in the continuous speech features, while retaining an autoregressive backbone to model longer-range dependencies.

\subsubsection{Patch-wise factorization.}
Given a continuous token sequence $\mathbf{x}=(x_1,\dots,x_N)$, the canonical autoregressive factorization is
\begin{equation}
p_\theta(\mathbf{x})=\prod_{i=1}^{N} p_\theta(x_i \mid x_{<i}).
\label{eq:ar_factorization}
\end{equation}
Rather than parameterizing Eq.~\eqref{eq:ar_factorization} directly at the frame level, we group neighboring tokens into patches of size $P$.
Let $\mathbf{x}^{(k)}=(x_{(k-1)P+1},\dots,x_{kP})$ denote the $k$-th patch.
Each patch is mapped by an aggregation encoder into an embedding, and the sequence of patch embeddings is processed by the causal autoregressive transformer.
The autoregressive transformer outputs a hidden representation $h_k$ summarizing the historical context up to patch $k$, which is used as the condition to generate the next patch:
\begin{equation}
p(\mathbf{x}) \approx \prod_{k} p_{\theta_a}(h_k \mid \mathbf{x}^{(\le k)}) \, p_{\theta_b}(\mathbf{x}^{(k+1)} \mid h_k),
\label{eq:patch_factorization}
\end{equation}
where $\theta_a$ denotes the causal autoregressive backbone and $\theta_b$ denotes the local diffusion decoder.

\subsubsection{Local diffusion decoding (next-patch generation).}
Following this formulation, the diffusion decoder (LocDiT in the original DiTAR design) performs \emph{next-patch generation} rather than full-sequence diffusion.
Concretely, after the autoregressive backbone produces $h_k$, the diffusion module takes $h_k$ as condition and denoises the target patch with bidirectional attention, enabling stronger modeling of intra-patch structure.
In addition, historical patches can be provided as prefix context to the diffusion decoder, turning local prediction into a context-aware ``outpainting''-style generation process and improving cross-patch continuity.

\subsubsection{Diffusion parameterization and training objective.}
The diffusion process over each target patch is instantiated as a variance-preserving (VP) forward process:
\begin{equation}
x_t = \alpha_t x_0 + \sigma_t \epsilon, \quad \epsilon \sim \mathcal{N}(0, I), \quad t \in [0,1],
\label{eq:vp_forward}
\end{equation}
and the diffusion decoder is trained with a conditional flow-matching objective:
\begin{equation}
\mathcal{L}_{\mathrm{diff}}
=
\mathbb{E}\left[
\left\lVert
v_\theta(x_t, t) - v(x_t, t)
\right\rVert_2^2
\right].
\label{eq:cfm_loss}
\end{equation}
where $v_\theta$ is the predicted velocity field and $v(x_t,t)$ is the target velocity induced by the forward process.

\noindent
In our work, we inherit this diffusion--autoregressive factorization as the generative backbone, and adapt the conditioning and training objectives to our task.

\subsection{Controllable Attributes}
\subsubsection{Coarse-grained conditions}
We incorporate \textbf{speaker identity} as a coarse-grained controllable attribute to guide the generation. Specifically, this can be achieved in two ways: by providing a speaker embedding or by providing prompt speech. This attribute serves as a global condition for each utterance, aiming to control timbre-related characteristics without requiring fine-grained frame-level annotations.\\

\noindent\textbf{Timbre} 
To capture and replicate the target speaker's unique vocal characteristics, the model takes as input a speaker embedding vector extracted from the prompt speech. We follow the approach of CosyVoice~\cite{du2024cosyvoice, du2024cosyvoice2} by using a pre-trained voiceprint model\footnote{\url{https://www.modelscope.cn/models/iic/CosyVoice-300M/file/view/master/campplus.onnx}} to extract the robust speaker embedding. This global vector ensures that the synthesized audio maintains high speaker similarity with the reference.\\


\subsubsection{Fine-grained conditions}
\textbf{Pitch (Fundamental Frequency)} \quad
Pitch corresponds to the fundamental frequency ($f_0$), which reflects the vocal fold vibration rate in voiced speech and plays an important role in prosody. In this work, $f_0$ is extracted using the WORLD vocoder~\cite{morise2016world}, where the DIO algorithm~\cite{morise2009fast} first estimates the raw contour and StoneMask is then applied for refinement. The refined $f_0$ is converted from Hz to the Mel scale to better match human auditory perception:
\begin{equation}
    f_{\text{mel}} = 1127 \ln\left(1 + \frac{f_0}{700}\right)
\end{equation}

The Mel-scaled $f_0$ is then quantized into \textbf{128 bins} (0--127) within a predefined range of 65.0--650.0 Hz. Voiced frames are linearly mapped to bins 1--126, unvoiced frames ($f_0 \le 0$) are assigned to bin 0, and values above 650.0 Hz are clipped to bin 127. The final coarse pitch is obtained by rounding to the nearest integer. \\

\noindent\textbf{Loudness} \quad
Loudness reflects the perceived intensity of sound. Since human hearing is more sensitive to mid-frequency components than to very low or high frequencies, the waveform is first processed with an A-weighting filter~\cite{IEC61672-1-2013} to approximate human auditory sensitivity. Frame-wise RMS energy is then computed and converted to the decibel scale:
\begin{equation}
    L_{\text{dB}} = 20 \log_{10}(\max(\text{RMS}, \epsilon))
\end{equation}
where $\epsilon = 10^{-10}$ prevents numerical issues during silent segments.

The resulting dB values are clipped to the range of -60.0 to 0.0 dB and quantized into \textbf{64 bins} (0--63) after min-max normalization. This provides a compact representation of perceptual loudness with approximately a 1 dB resolution.\\

\noindent\textbf{Duration} \quad We specify the duration as the temporal length of each phone, which provides the model low-level control for speech pacing and rhythm. We use forced alignment to temporally align the phonetic transcription of each utterance with its paired waveform, and then specify the duration for each phone in terms of the total number of acoustic frames spanning that phone.

\subsection{\modelname}
\modelname\, builds on DiTAR~\cite{jia2025ditar} and extends it with explicit controllability for expressive speech synthesis. Given an input phone sequence, we first encode the linguistic content with phone embeddings. To enable fine-grained prosody control, each phone embedding is further augmented with aligned control signals, including discretized pitch, loudness, and duration. In addition, a speaker embedding extracted from prompt speech is used as a global condition to preserve the target speaker's timbre.

For acoustic modeling, we represent the target waveform as a sequence of continuous latent tokens produced by the VAE codec. Following the patch-wise design of DiTAR, we group the acoustic latent sequence into non-overlapping patches, where each patch contains 4 consecutive latent tokens. This patch size provides a simple and effective trade-off between local acoustic coherence and modeling efficiency. The resulting patch sequence is then fed into the autoregressive backbone for long-range dependency modeling.

Specifically, the concatenation of text/control tokens and acoustic patch tokens is processed by a transformer backbone, using modality-aware position indexing that resets positions independently for the text and speech streams. The autoregressive backbone outputs one context representation for each acoustic patch, which conditions the local diffusion transformer decoder to reconstruct the next patch in parallel. The entire model is optimized with an L1 flow-matching objective, together with an auxiliary stop prediction loss that classifies each acoustic patch as first, middle, or last~\cite{he2025continuous}. The final training objective is defined as
\begin{equation}
\mathcal{L} = \mathcal{L}_{\mathrm{flow}} + \lambda \mathcal{L}_{\mathrm{stop}},
\end{equation}
where $\mathcal{L}_{\mathrm{flow}}$ denotes the flow-matching loss, $\mathcal{L}_{\mathrm{stop}}$ denotes the stop prediction loss, and $\lambda$ is a balancing coefficient.

\section{Experimental Setup}
\subsection{Datasets}
During the pretraining stage, \modelname{} is trained on a subset of the English split of Emilia~\cite{he2024emilia} and a subset of GigaSpeech~\cite{chen2021gigaspeech}, yielding approximately 20,000 hours of English speech in total. This pretraining corpus provides broad acoustic and linguistic coverage, enabling the model to acquire robust speech generation capabilities. For evaluation, we assess zero-shot synthesis on LibriSpeech-PC \textit{test-clean}~\cite{chen2025f5} and Seed-TTS \textit{test-en}~\cite{anastassiou2024seed}, and evaluate fine-grained controllability on LJSpeech~\cite{ito2017lj}.


\subsection{Model Configuration}
\label{sec:model_config}
We train two model variants, a 0.1B model and a 0.6B model, under the same DiTAR architecture. The 0.1B model uses a hidden size of 512, with a 4-layer, 8-head aggregation encoder and a 4-layer, 8-head DiT decoder; in contrast, the 0.6B model scales the hidden size to 1024, and correspondingly adopts a 6-layer, 16-head aggregation encoder and a 6-layer, 16-head DiT decoder. For VAE, we directly use checkpoints from Semantic-VAE~\cite{niu2025semantic}.

\subsection{Training and Inference}
Both model variants were trained with the same setup. We used the AdamW optimizer with a learning rate of $2 \times 10^{-4}$ and a weight decay of 0.01. The learning rate was linearly warmed up over the first 5\% of training steps and then linearly decayed for the remaining steps. Each model was trained for 5 epochs on 8 NVIDIA A100 80GB GPUs. During inference, classifier-free guidance (CFG)~\cite{ho2022classifier} is adopted with 32 sampling steps and a guidance scale of 1.5, where the time and speaker embeddings are treated as a unified conditioning signal.

\subsection{Evaluation Metrics}
\textbf{Objective metrics}
We use Word Error Rate (WER) as an automatic proxy for the intelligibility of the synthesized speech; this is calculated using Whisper-large-v3~\cite{radford2023robust}. Additionally, speaker similarity (SIM-o) is objectively measured by computing the cosine similarity of speaker embeddings, which are extracted from both the generated and original target speech using a WavLM-based speaker verification model~\cite{chen2022wavlm}. RMSE for pitch and loudness measures frame-level deviation from the ground-truth speech in Hz and dB, respectively. For duration, we report phoneme-level mean absolute error (MAE). Lower RMSE or MAE indicates better controllability.

\noindent\textbf{Subjective metrics}
We adopt Comparative Mean Opinion Score (CMOS) and Similarity Mean Opinion Score (SMOS) assessed by human raters. For CMOS, raters listen to the synthesized utterance and the ground-truth recording and judge which one sounds more natural (and by how much). For SMOS, raters score the perceived speaker similarity between the synthesized speech and the reference prompt.

\section{Results and Analysis}
\label{sec:results}
\begin{table}[htb]
\scriptsize
\centering
\caption{Zero-Shot TTS performance on LibriSpeech-PC test-clean and Seed-TTS test-en.}
\label{tab:main}
\begin{tabular}{lcccc}
\toprule
\textbf{Model} & \textbf{WER(\%)$\downarrow$} & \textbf{SIM-o$\uparrow$} & \textbf{CMOS$\uparrow$} & \textbf{SMOS$\uparrow$} \\
\midrule

\multicolumn{5}{c}{\textbf{LibriSpeech-PC \textit{test-clean}}} \\
\midrule
Ground Truth & 2.40 & 0.69 & 0.00 & 3.90 \\
Vocoder Reconstructed & 2.45 & 0.68 & - & - \\
\hdashline
DiTAR$^\dagger$ & 2.55 & 0.61 & -0.22 & 3.74 \\
CtrlSpeech (0.1B) & 4.36 & 0.61 & -0.65 & 3.66 \\
CtrlSpeech (0.6B) & \textbf{2.46} & \textbf{0.65} & -0.16 & 3.81\\

\midrule

\multicolumn{5}{c}{\textbf{Seed-TTS \textit{test-en}}} \\
\midrule
Ground Truth & 1.90 & 0.73 & 0.00 & 3.85 \\
Vocoder Reconstructed & 1.93 & 0.69 & - & - \\
\hdashline
DiTAR$^\dagger$ & 2.89 & 0.59 & -0.32 & 3.70 \\
CtrlSpeech (0.1B) & 6.50 & 0.58 & -0.60 & 3.66 \\
CtrlSpeech (0.6B) & \textbf{2.58} & \textbf{0.63} & -0.23 & 3.78 \\
\bottomrule
\end{tabular}
\parbox{\linewidth}{\vspace{0.5ex}\footnotesize
\raggedright $^\dagger$Since DiTAR is not open-source, these results are based on our own reproduction using the same training data. The model size is 0.6B.}
\end{table}
\subsection{Zero-shot TTS Performance}
Table~\ref{tab:main} summarizes the zero-shot TTS results on LibriSpeech-PC \textit{test-clean} and Seed-TTS \textit{test-en}. Overall, \modelname{} (0.6B) achieves competitive performance against the reproduced DiTAR baseline. On LibriSpeech-PC, it obtains 2.46\% WER and 0.65 SIM-o, and on Seed-TTS, it reaches 2.58\% WER and 0.63 SIM-o, outperforming DiTAR on both objective metrics. Subjective results also show consistent gains in SMOS, indicating better speaker similarity in zero-shot synthesis.

\subsection{Effect of Speaker Control}
\begin{table}[htb]
\footnotesize
\centering
\caption{Ablation study on coarse-grained speaker control conditions for \modelname{} (0.6B) on the Seed-TTS test-en set.}\label{tab:speaker_control_ablation}
\begin{tabular}{lcc}
\toprule
\textbf{Condition} & \textbf{WER(\%)$\downarrow$} & \textbf{SIM-o$\uparrow$} \\
\midrule
w/o speaker condition & 3.12 & 0.07 \\
+ speaker embedding only & 3.27 & 0.48 \\
+ prompt speech only & 2.69 & 0.53 \\
+ embedding + prompt speech & \textbf{2.58} &  \textbf{0.63} \\
\bottomrule
\end{tabular}
\end{table}
Table~\ref{tab:speaker_control_ablation} shows the effect of coarse-grained speaker conditions. Without speaker conditioning, SIM-o drops to 0.07, confirming its importance for timbre control. Using only speaker embedding or only prompt speech improves performance, while combining both achieves the best result (2.58\% WER and 0.63 SIM-o), indicating that they provide complementary speaker information.

\subsection{Fine-grained Controllability}
\begin{table}[htb]
\scriptsize
\caption{RMSE between synthesized speech and ground truth speech on LJSpeech test set.}
\label{tab:pitch-loudness}
\centering
\setlength{\tabcolsep}{6pt}
\renewcommand{\arraystretch}{1.12}
\begin{tabular}{@{}llcc@{}}
\toprule
\textbf{Condition} & \textbf{Model} & \textbf{Pitch (Hz)$\downarrow$} & \textbf{Loudness (dB)$\downarrow$} \\
\midrule
\multirow{3}{*}{Text-only}
& DrawSpeech~\cite{chen2025drawspeech}         & 43.59 & 14.03 \\
& CtrlSpeech (0.1B)  & 77.84 & 7.12 \\
& CtrlSpeech (0.6B)  & 67.86 & 6.35 \\
\addlinespace[2pt]
\hline
With Sketch & DrawSpeech~\cite{chen2025drawspeech} & \textbf{27.78} & 13.15 \\
\multirow{2}{*}{With Control Signals} & CtrlSpeech (0.1B)  & 41.54 & 5.01 \\
& CtrlSpeech (0.6B)  & 38.39 & \textbf{4.56} \\
\bottomrule
\end{tabular}
\end{table}
As shown in Table~\ref{tab:pitch-loudness}, we first measure pitch and loudness controllability on the LJSpeech test set using RMSE with respect to the ground-truth speech. In the text-only setting, \modelname{} exhibits larger pitch error than DrawSpeech, indicating that accurate local prosody is difficult to recover from text alone. However, once explicit control signals are provided, both model variants show substantial reductions in RMSE for both attributes. In particular, CtrlSpeech (0.6B) reduces pitch RMSE from 67.86 Hz to 38.39 Hz and loudness RMSE from 6.35 dB to 4.56 dB, demonstrating that the aligned pitch and loudness conditions provide effective and direct control over local acoustic realization.

\begin{table}[htb]
\centering
\caption{Phoneme-level duration mean absolute error (MAE) on LibriSpeech-PC test-clean.}
\label{tab:duration}
\footnotesize
\setlength{\tabcolsep}{7pt}
\renewcommand{\arraystretch}{1.15}
\begin{tabular}{@{}lcc@{}}
\toprule
\textbf{Model} & \textbf{Setting} & \textbf{MAE $\downarrow$} \\
\midrule
DiTAR (replicated) & -- & 28.00 \\
\midrule
\multirow{2}{*}{CtrlSpeech (0.1B)}
& w/o control & 31.58 \\
& w/ control & 14.00 \\
\midrule
\multirow{2}{*}{CtrlSpeech (0.6B)}
& w/o control & 28.08 \\
& w/ control & \textbf{11.86} \\
\bottomrule
\end{tabular}
\end{table}
Table~\ref{tab:duration} further evaluates duration controllability using phoneme-level duration MAE on LibriSpeech-PC \textit{test-clean}. Without duration control, \modelname{} achieves similar duration accuracy to the replicated DiTAR baseline. In contrast, when duration control is enabled, the error drops sharply for both model sizes. Specifically, the MAE decreases from 31.58 to 14.00 for CtrlSpeech (0.1B), and from 28.08 to 11.86 for CtrlSpeech (0.6B), substantially outperforming the replicated DiTAR result. These results confirm that the proposed phone-aligned duration condition allows the model to better follow target temporal patterns and gives users effective control over speech pacing and rhythm.


\subsection{User Interface for Coarse-to-Fine Control}
As shown in Figure~\ref{fig:control-pipeline}, the user interface of \modelname{} supports a coarse-to-fine workflow: users first generate an initial speech sample with input text and speaker conditions, and then iteratively refine local pitch, loudness, and duration through fine-grained controls.

\section{Conclusion}
In this paper, we proposed \modelname{}, a controllable expressive TTS framework with coarse-to-fine control. By combining global speaker conditioning with phone-aligned prosodic signals, \modelname{} enables flexible control over expressive speech while preserving zero-shot synthesis quality. Experimental results demonstrate its effectiveness in zero-shot TTS, as well as its coarse- and fine-grained controllability.

\section{Limitations}

This work has several limitations. First, our experiments are mainly conducted on English speech, so the effectiveness of \modelname{} for multilingual or code-switching synthesis remains unexplored. Second, the proposed fine-grained control depends on phone-aligned pitch, loudness, and duration signals, whose quality may be affected by errors from pitch extraction and forced alignment. Third, although explicit controls improve attribute controllability, text-only generation still has difficulty predicting precise local prosody, especially pitch. Finally, the current system focuses on speaker identity, pitch, loudness, and duration, while other expressive factors such as emotion, voice quality, and nonverbal vocalizations are not explicitly modeled.

\section{Acknowledgments}
This work was supported by Amazon.com, PO No. 2D-16003984 through the Amazon-UT Austin HUB. We thank Guanrou Yang and Zhikang Niu for their incredible help.


\section{Generative AI Use Disclosure}
Generative AI tools were used only for language editing and proofreading. All authors remain fully responsible for the content of this manuscript.

\bibliographystyle{IEEEtran}
\bibliography{mybib}

\end{document}